\documentclass[aps,pra,reprint,superscriptaddress]{revtex4-2}

\usepackage{graphicx}
\usepackage{bm}
\usepackage{amsmath}
\usepackage{url}

\begin{document}
\title{Pattern Formation Simulated by an Ising Machine}

\author{Kanon Mukai}
\email[]{mukai.kanon@is.ocha.ac.jp}
\affiliation{Department of Computer Science, Ochanomizu University, Bunkyo,
Tokyo 112-8610, Japan}
\author{Kazue Kudo}
\email[]{kudo@is.ocha.ac.jp}
\affiliation{Department of Computer Science, Ochanomizu University, Bunkyo,
Tokyo 112-8610, Japan}
\affiliation{Graduate School of Information Sciences, Tohoku University,
Sendai 980-8579, Japan}

\begin{abstract}
 In a ferromagnetic Ising system, domain pattern formation, i.e., phase-ordering, occurs after a sudden quench.
We propose the method to simulate the pattern formation dynamics by an Ising machine.
We demonstrate that the method reproduces domain patterns similar to those simulated by the Monte Carlo method.
Moreover, the same domain growth law is observed in the proposed method and the Monte Carlo method.
\end{abstract}

\maketitle

Ising machines are special-purpose computers for solving combinatorial optimization problems and sampling.
Several hardware devices, as well as quantum annealers, have been developed recently~\cite{Johnson2011, Inagaki2016, Yamaoka2016, Aramon2019, Goto2019}, inspired by quantum annealing.
Accepting a Hamiltonian formulated by the Ising model or quadratic unconstrained binary optimization (QUBO) formulation, an Ising machine returns the ground state or low-energy states of the Hamiltonian.
Ising machines are usually used for solving combinatorial optimization problems, whose solutions are the ground states of the corresponding Hamiltonians.
We here focus on a method of computing the time evolution, specifically the dynamics of domain pattern formation, which may look unrelated to ground states.

When an Ising system is suddenly quenched from the disordered phase to the ferromagnetic phase, small clusters are formed. 
They merge to form larger domains and grow with time $t$.
The domain growth dynamics show scaling behavior, which can be captured by rescaling length using the characteristic length scale $L(t)$.
For instance, the enlarged part of a domain pattern at an early time looks similar to a domain pattern at a later time.
In a nonconserved two-dimensional Ising system, the characteristic length grows as $L(t)\sim t^{1/2}$ in the late stage of growth~\cite{Bray1994}.

In this paper, we propose the method to simulate the pattern formation dynamics of the Ising model by using an Ising machine.
If we use an Ising machine in a usual manner, we only obtain the ground state of the Ising model, i.e., the final state of a domain pattern.
Thus, we need another method to simulate the dynamics of domain pattern formation.
The essential idea is to incorporate the last spin orientation at each site as a local field.
The model can be derived based on the time-dependent Ginzburg-Landau (TDGL) equation.
We demonstrate domain patterns and the domain growth dynamics simulated by both the Monte Carlo method and the method using an Ising machine.

We simulate pattern formation of the nonconserved Ising model after a sudden quench from a high temperature to a temperature below the critical temperature.
The Hamiltonian of the model is written as
\begin{equation}
H=-J\sum_{\langle i,j\rangle} S_iS_j,
\label{eq:H.0}
\end{equation}
where $J>0$ and $S_i=\pm 1$ denotes spin at site $i$, 
and the sum is taken over all nearest-neighbor pairs $\langle i,j\rangle$. 
The simple way to perform the simulation is to employ the single spin-flip Monte Carlo method, which leads to the Glauber dynamics.

Here, we derive the model for simulating the pattern formation using an Ising machine.
The Ising model Hamiltonian corresponds to the Ginzburg-Landau free energy, which is written as, in two-dimension,
\begin{equation}
F=\int\! d^2\bm{r}\left[ f(m) +\frac{D}{2}\left(\bm\nabla m\right)^2\right],
\label{eq:F.0}
\end{equation}
where $m=m(\bm{r},t)$ is the magnetization, 
$D$ is the constant that corresponds to the exchange interaction, and
\begin{equation}
f(m) = \frac14 m^4 - \frac12 m^2
+ \textrm{const.}
\end{equation}
Then, the TDGL equation with thermal noise is given by
\begin{equation}
 \frac{\partial}{\partial t} m(\bm{r},t)
= -\left[ f'(m) -D \nabla^2 m \right] + \eta(\bm{r},t).
\label{eq:TDGL}
\end{equation}
Here, $\eta(\bm{r},t)$ is the Gaussian random noise with zero mean and satisfies
$\langle\eta(\bm{r},t)\eta(\bm{r}',t')\rangle = 2k_BT\delta(\bm{r}-\bm{r}')\delta(t-t')$, where $k_B$ is the Boltzmann constant, and $T$ is temperature.
Replacing the time derivative with the backward difference, we have
\begin{equation}
\frac{m^k-m^{k-1}}{\Delta t}= -
\left[ f'(m^k) -D \nabla^2 m^k \right] + \eta^k,
\label{eq:diff}
\end{equation}
where $m^k=m(\bm{r}, k\Delta t)$ with $k$ being an integer, and $\Delta t>0$.
We take the noise term as
\begin{align}
 \eta^k=\frac{1}{\Delta t}\int_{k\Delta t}^{(k+1)\Delta t}
\eta(\bm{r}, t) dt.
\end{align}
Here, $\eta^k$ corresponds to a random number with zero mean, and its variance is $2k_BT/\Delta t$.

We here introduce another free energy
\begin{equation}
 \mathcal{F} =  F^k 
- \frac{1}{\Delta t}\int\! d^2\bm{r}\; m^{k-1}m^k 
+ \frac{1}{2\Delta t}\int\! d^2\bm{r}\; (m^k)^2
- \int\! d^2\bm{r}\;  \eta^k m^k,
\label{eq:F.1}
\end{equation}
where $F^k$ is given by Eq.~\eqref{eq:F.0} at $t=k\Delta t$. 
From the condition $\delta\mathcal{F}/\delta m^t=0$, which minimizes $\mathcal{F}$, we obtain the same equation as Eq.~\eqref{eq:diff}.
Therefore, minimizing $\mathcal{F}$ at each time step corresponds to solving the TDGL equation under a first-order approximation.

Now, we discretize Eq.~\eqref{eq:F.1}. 
Then, $m^k$ corresponds to $S_i^k=\pm 1$.
Under the approximation $(m^k)^2=1$, 
$f(m^k)$ in Eq.~\eqref{eq:F.0} is constant.
The second term of the right-hand side of Eq.~\eqref{eq:F.0} corresponds to
\begin{align}
 \frac{D}{2}\sum_{\langle i,j\rangle}
\left(\frac{S_i^k-S_j^k}{\Delta x}\right)^2 
=-\frac{D}{(\Delta x)^2}\sum_{\langle i,j\rangle}S_i^k S_j^k
+\textrm{const.}
\end{align}
Here, $\Delta x$ is the lattice constant.
Ignoring the constant terms, we have the effective model,
\begin{align}
\mathcal{H} &= -J\sum_{\langle i, j\rangle} S_i^k S_j^k - \sum_i h_i^k S_i^k,
\label{eq:H.1}\\
h_i^k &= KS_i^{k-1} + \eta_i^k,
\label{eq:h}
\end{align}
where $J=D/(\Delta x)^2$ and $K=1/\Delta t$.

The algorithm to simulate pattern formation using an Ising machine, which we call the annealing method, is as follows.
The initial state is set as $S_i^0=\pm 1$ at random.
Each time, the local filed $h_i^k$ is given by Eq.~\eqref{eq:h}, where $\eta_i^k$ is a Gaussian random number.
Using an Ising machine, we obtain the spin configuration minimizing Eq.~\eqref{eq:H.1}.
The configuration is used for the local field at the next time step.

The simulations are performed in a $N\times N$ square lattice with $N=256$.
The boundary condition is periodic for the Monte Carlo method and is open for the annealing method.
We fix $J=1$ both in Eqs.~\eqref{eq:H.0} and \eqref{eq:H.1}.
The inverse temperature $\beta=1/(k_BT)$ corresponds to thermal noise $\eta_i^k$, which is given by the Gaussian random number with zero mean and variance of $\sigma^2=2k_BT/\Delta t=2K/\beta$ in the annealing method.
As an Ising machine, we use the CMOS annealing machine (GPU version) provided by Annealing Cloud Web~\cite{ACW}.

\begin{figure}[tb]
\centering
\includegraphics[width=8cm]{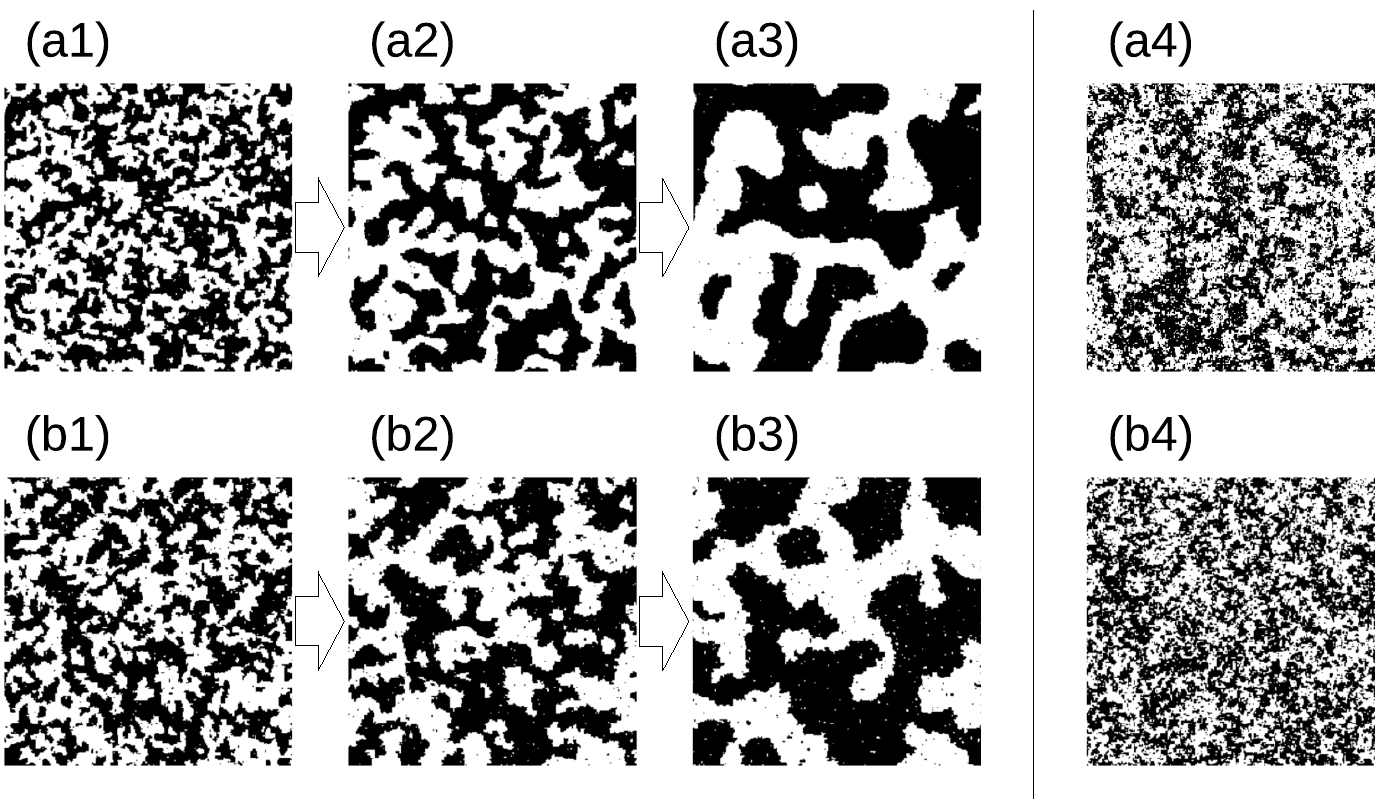}
\caption{Snapshots of domain patterns simulated by (a) the Monte Carlo and (b) the annealing methods. 
Domains grow for $\beta=0.8$ (below the critical temperature) in both the Monte Carlo method [(a1) 10 MCS, (a2) 40 MCS, (a3) 200 MCS] and the annealing method [(b1) 15 steps, (b2) 60 steps, (b3) 300 steps].
However, no ordered patterns appear for $\beta=0.4$ (above the critical temperature) in either the Monte Carlo method [(a4) 200 MCS] or the annealing method [(b4) 300 steps].
The inverse timestep is $K=2$ for (b1--b4).
}
\label{fig:snapshots}
\end{figure}

Snapshots in Fig.~\ref{fig:snapshots} demonstrate the time evolution of pattern formation that is simulated by (a1--a3) the Monte Carlo method and (b1--b3) the annealing method.
The time unit is one Monte Carlo step (MCS) per spin, which corresponds to $N^2$ spin-flip attempts, for the Monte Carlo method.
In the annealing method, one step is the update of spin configuration using the Ising machine.
The inverse temperature is set as $\beta=0.8$, which corresponds to $\sigma^2=5$ for (b1--b3) and is below the critical temperature.
We see that domains grow with time in both the Monte Carlo and the annealing methods.

In contrast, no ordered domain patterns appear above the critical temperature, as shown in Figs.~\ref{fig:snapshots}(a4) and \ref{fig:snapshots}(b4), which are at 200 MCS and 300 steps, respectively.
The inverse temperature is set as $\beta=0.4$, which corresponds to $\sigma^2=10$ for (b4).

\begin{figure}[tb]
\centering
\includegraphics[width=8cm]{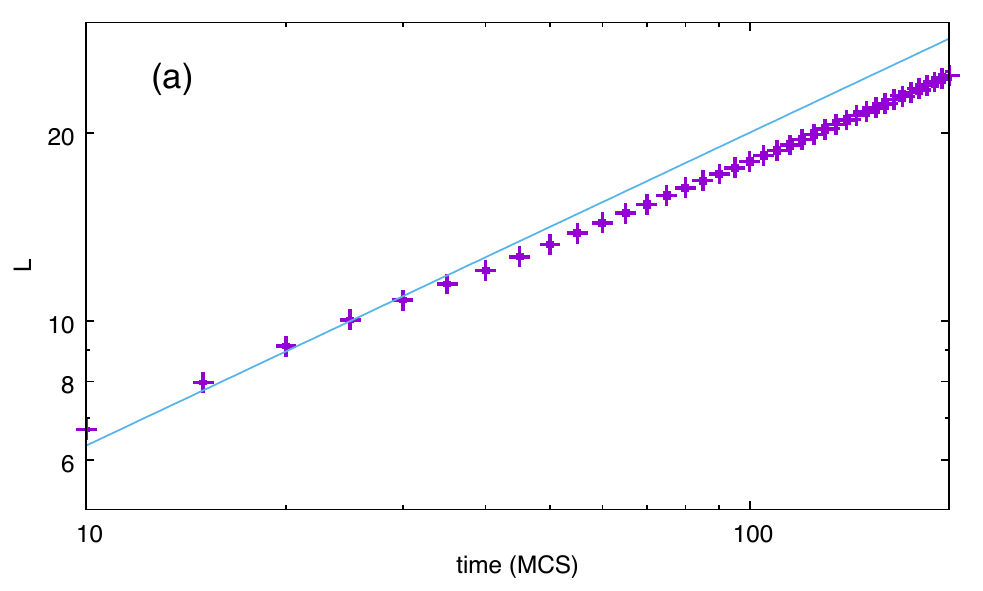}
\includegraphics[width=8cm]{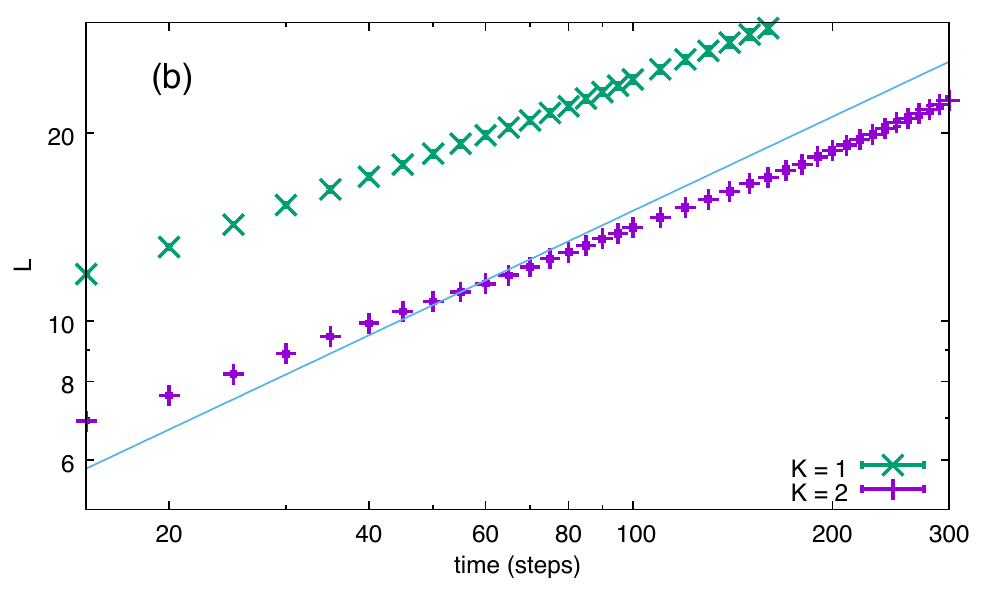}
\caption{(Color online) 
Time evolution of the characteristic length $L(t)$ for (a) the Monte Carlo method and (b) the annealing method.
Solid lines represent $L(t)\propto t^{1/2}$.
Error bars, which are too small to recognize, represent the standard error.}
\label{fig:length}
\end{figure}

Figure~\ref{fig:length} illustrates the time evolution of the characteristic length $L(t)$ of domain patterns for $\beta=0.8$.
Here, we define $L(t)$ as
\begin{equation}
C(L(t)/2,t) = 0.5C(0,t),
\label{eq:Lt}
\end{equation}
where $C(r,t)$ is the correlation function, which is defined by
\begin{equation}
C(r,t) = \left\langle \frac{1}{A}
\int d^2\bm{r}' S(\bm{r}',t)S(\bm{r}+\bm{r}',t)\right\rangle.
\label{eq:Cr}
\end{equation}
Here, $A$ is the area of the system, and $\langle\cdots\rangle$ represents the azimuth average.
Since the system is discrete actually, $S(\bm{r},t)$ corresponds to $S_i^k$, and $A=N^2$.
Each point in Fig.~\ref{fig:length} represents the average taken over 20 simulations.
The characteristic length grows as $t^{1/2}$ approximately for both the Monte Carlo and the annealing methods in the long-time regime. 
In Fig.~\ref{fig:length}(b), since the timestep width for $K=1$ is larger than that for $K=2$, the characteristic length is larger for $K=1$ than $K=2$ after the same number of steps.

In conclusion, the proposed method, referred to as the annealing method, simulates the pattern formation of the Ising model and well reproduces the pattern formation dynamics simulated by the Monte Carlo method.
Actually, the annealing method is not efficient for the pattern formation simulation in a simple ferromagnetic Ising system.
This work is the starting point to apply the method to other systems, such as a system with long-range interactions.
The annealing method is a method mapping dynamics to a variational problem.
Thus, the method can be applied to other types of problems if they are mapped to variational problems.

\begin{acknowledgments}
The authors thank K.~Hukushima for the fruitful discussion.
This work is partially supported by JSPS KAKENHI Grant Number JP18K11333.
\end{acknowledgments}


\end{document}